\documentclass{article}

\usepackage[final,nonatbib]{styles/tackling_climate_workshop_style}

\usepackage[utf8]{inputenc} 
\usepackage[T1]{fontenc}    
\usepackage{url}            
\usepackage{booktabs}       
\usepackage{amsfonts}       
\usepackage{nicefrac}       
\usepackage{microtype}      
\usepackage{graphicx}
\usepackage{amsmath}
\usepackage{xcolor}

\title{Emulating the Global Change Analysis Model\\with Deep Learning}

\author{%
  Andrew Holmes$^1$, Matt Jensen$^2$, Sarah Coffland$^1$, Hidemi Mitani~Shen$^1$, Logan Sizemore$^1$,\\Seth Bassetti$^3$, Brenna Nieva$^1$, Claudia Tebaldi$^4$, Abigail Snyder$^4$, Brian Hutchinson$^{1,4}$ \\
  $^1$ Computer Science Dept, Western Washington University, Bellingham, WA, USA\\
  $^2$ Applied Artificial Intelligence Systems, Pacific Northwest National Laboratory, Seattle, WA, USA\\
  $^3$ Computer Science Department, Utah State University, Logan, UT, USA\\
  $^3$ Utah State University, Logan, UT, USA\\
  $^4$ Joint Global Change Research Institute, Pacific Northwest National Laboratory, College Park, MD\\
}

\begin{document}

\maketitle

\begin{abstract}
  The Global Change Analysis Model (GCAM) simulates complex interactions between the coupled Earth and human systems, providing valuable insights into the co-evolution of land, water, and energy sectors under different future scenarios. Understanding the sensitivities and drivers of this multisectoral system can lead to more robust understanding of the different pathways to particular 
  outcomes. The interactions and complexity of the coupled human-Earth systems make GCAM simulations costly to run at scale - a requirement for large ensemble experiments which explore uncertainty in model parameters and outputs. A differentiable emulator with similar predictive power, but greater efficiency, could provide novel scenario discovery and analysis of GCAM and its outputs, requiring fewer runs of GCAM. As a first use case, we train a neural network on an existing large ensemble that explores a range of GCAM inputs related to different relative contributions of energy production sources, with a focus on wind and solar. We complement this existing ensemble with interpolated input values and a wider selection of outputs, predicting $22,528$ GCAM outputs across time, sectors, and regions. 
  We report a median $R^2$ score of $0.998$ for the emulator's predictions and an $R^2$ score of $0.812$ for its input-output sensitivity. 
\end{abstract}

\section{Introduction and Background}
\label{sec:intro}

The global change problem involves both Earth and human system dynamics, interacting and creating feedbacks among the multiple components and sectors that make up the whole system.
 The Global Change Analysis Model (GCAM) \cite{bond-lambertyGcamdataPackagePreparation2019,calvinGCAMV5Representing2019} and other models of the same class are essential to represent the future evolution of the human system, including socioeconomic, land, energy, and water sectors, giving rise to future plausible and coherent scenarios of emissions. These scenarios are in turn used as drivers of Earth system model projections. In the opposite direction, climate output from Earth system models is used to model impacts in GCAM and other integrated multi-sector models. This work focuses on emulating GCAM specifically; it is an open-source multisector dynamic model that simulates the integrated, simultaneous evolution of energy, agriculture, land use, water, and climate system components.
GCAM simulates global markets segmented into 32 distinct socioeconomic regions, 235 hydrological basins, forming 384 land units from the intersection of basins and regions.

Historically, GCAM and comparable models have run a discrete set of ``storylines'' or representative future scenarios. In contrast, thanks to advances in computational power and analysis tools, exploratory modeling, sampling a much larger set of drivers (and therefore outcomes) has become popular in recent years \cite{dolanModelingEconomicEnvironmental2022a, dolanEvaluatingEconomicImpact2021a}. 
In this approach, large ensembles of scenarios are designed and run to fill the gaps between the representative storyline scenarios.
This approach has been fruitful for exploring the complex sensitivities these models have to assumptions about the systems under test, and the external drivers that determine their outcomes. This understanding of sensitivity and drivers can facilitate identification of pros and cons of different pathways to outcomes of interest (e.g., to minimize water scarcity \cite{dolanEvaluatingEconomicImpact2021a}).
The ensembles are often designed to incorporate a range of data sources, expert opinion, and discrete parameterizations in a factorial combination~\cite{woodardScenarioDiscoveryAnalysis2023}. However, even with access to modern computing clusters, computational cost hinders a comprehensive exploration of these inputs. 
We aim to enable this comprehensive exploration via deep learning-based emulation of GCAM. 
Existing large ensembles provide data 
 to train and evaluate such emulators.

Once trained, a high-fidelity emulator can be used to aid our understanding both of the coupled Earth-human systems and their models (e.g., GCAM). For example, an emulator could be used to explore the input (assumption) space, to steer the generation of large GCAM ensembles, or to better characterize model sensitivities. 
There are two defining aspects to our approach that set it apart from GCAM toward these goals. First, once trained, predicting outcomes for novel scenarios is faster than GCAM by at least three orders of magnitude. Second, the differentiability of the emulation enables efficient search algorithms over the input space.
Relatively little work has been done with emulation of integrated, multisector models, but results have been promising \cite{takakuraReproducingComplexSimulations2021,xiongEmIAMV1Emulator2023}.
Here we introduce a high-fidelity emulator of GCAM, both in the predictions and in the input-output sensitivities.

\section{Methods}

\subsection{Data}
Each scenario of GCAM is shaped by exogenous factors like socioeconomic trends (population and GDP growth), technology costs and performance, historical information, and assumptions about future values of key drivers. These are what we call ``inputs'' in this  paper, and a subset of these will be sampled in our ensembles. 
GCAM provides a detailed, time-evolving analysis of sectors within the economy and simulates how different external factors might affect specific sectors over time, taking into account the effects from all other sectors; these serve as the outputs of our emulator.

\paragraph{Inputs:}
\label{sec:inputs}

This paper follows the experiment set up by Woodard et al.~\cite{woodardScenarioDiscoveryAnalysis2023}, W2023 henceforth, to study the effect of varying inputs on wind and solar energy adoption by 2050. We use the same 12 GCAM inputs as W2023, representing
costs, constraints, backups, and demand in the energy sector.
These factors were chosen by climate experts to describe a wide variety of scenarios to explore GCAM and its outputs.  Table \ref{tab:inputs} describes each of the 12 inputs. In W2023 experiments, these factors were held to \emph{high} and \emph{low} values which were encoded as 1 and 0, respectively, in our experiments. 

To enrich the input space, we consider here input values between the high and low. For nine of the 12 inputs (see Appendix~\ref{app:inputs}), an \emph{intermediate} value between high and low is well-defined, so we relax the domain from $\{0, 1\}$ to the interval $0 \leq x \leq 1$. The extreme high and low values still represent the original binary meaning, while all intermediate values are linearly interpolated between the high and low scenarios. For three of the twelve inputs, a notion of \emph{intermediate} is not well-defined; namely, for bioenergy, electrification, and emissions, the binary values represent the presence of absence of specific input files to GCAM. 

\paragraph{Sampling Strategies:}
\label{sec:sampling}
With the introduction of interpolated values, the input space can no longer be enumerated, so we consider two strategies for sampling the space: Latin hypercube~\cite{mckayComparisonThreeMethods1979}  and ``finite-diff'' \cite{sobolGlobalSensitivityIndices2001,sobolDerivativeBasedGlobal2009}. In either strategy, the nine interpolated inputs are sampled by the strategy while the remaining three inputs are randomly sampled randomly uniformly in $\{0,1\}$. We selected Latin hypercube  
to efficiently explore the interpolated input space, while the finite-diff was selected to support our sensitivity analysis.
We sample 4096 input configurations for Latin hypercube data (denoted here ``interpolated'') data, which we split into training, validation and test sets at an 80\%/10\%/10\% ratio. The finite-diff data (denoted here as the ``DGSM'' or ``sensitivity'' dataset) contains 4000 samples and is entirely test set, as it was used neither for model training nor tuning. 

\paragraph{Outputs:}
\label{sec:outputs}

Each GCAM run produces a large output database related to the energy, water, climate, and land sectors. Among these, we identify 44 GCAM output quantities to predict (see Appendix~\ref{app:output-quantities} for full details). These quantities were chosen to cover physical quantities and prices over the major resources in the water, land, and energy sectors relevant to renewable energy adoption, in light of the focus in W2023. For each of the 44 output quantities, GCAM and our emulator predict values over 32 regions and over 16 model years, $\{ 2025, 2030, 2035, \dots, 2095, 2100\}$. This yields a total output dimension of $22,528$  values to predict.

\subsection{Emulator}
\label{sec:emulator}

\begin{figure}
    \centering
    \includegraphics[width=\linewidth]{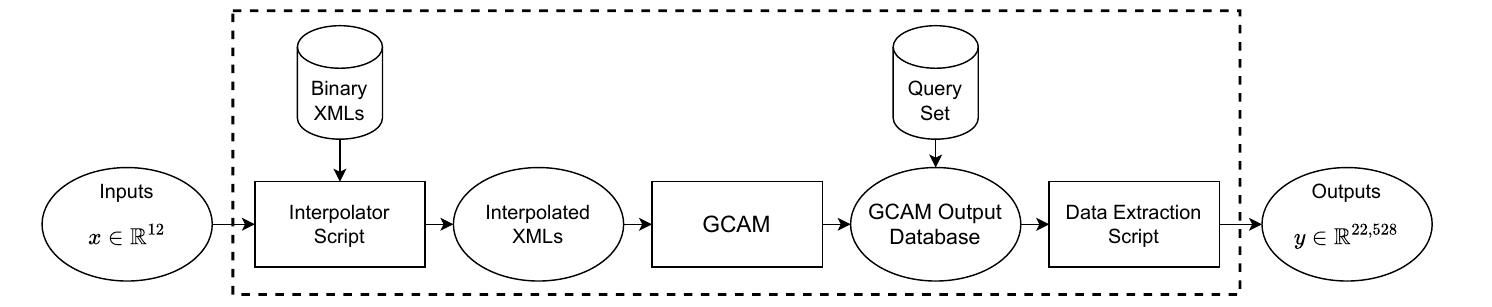}
    \caption{Diagram of the input-output relationship using GCAM. The emulator approximates the dashed box, mapping
directly from inputs to outputs}
    \label{fig:gcam-overview}
\end{figure}

Figure~\ref{fig:gcam-overview} illustrates the emulation problem. Our emulator abstracts a series of steps between inputs and outputs, including interpolating the configuration XMLs, running GCAM, and running queries to extract the output values of interest.

\paragraph{Model Architecture:}
Motivated by the success of neural networks learning non-linear relationships \cite{goodfellowDeepLearning2016}, we employ a nueral network to emulate input-output relationship (dashed box of Fig.~\ref{fig:gcam-overview}). 
Specifically, we use a four-layer, feed-forward, fully connected neural network, each layer with 256 hidden units followed by a linear rectified unit (ReLU) hidden activation function \cite{fukushimaVisualFeatureExtraction1969}. The fully connected output layer contains $22,528$ units. 

\paragraph{Training:}
The model is trained to minimize mean squared error loss between the emulator predictions and GCAM outputs on the training set using hyperparameters selected with a Bayesian Hyperparameter search \cite{snoek2012practical} via Weights and Biases \cite{wandb} on the validation set. All output values are z-score normalized using their training set statistics -- each quantity-region-year pair $x_{qry}$ is normalized using $(x_{qry} - \mu_{qry}) / \sigma_{qry}$; where mean $(\mu_{qry})$ and standard deviation $(\sigma_{qry})$ are computed for that specific quantity-region-year value across all training dataset scenarios. We train with the AdamW \cite{loshchilovDecoupledWeightDecay2017} stochastic optimization algorithm for $500$ epochs with a learning rate of $0.001$. 

\section{Results and Analysis}
\label{sec:results}

\begin{table}[b]
    \centering
    \caption{Evaluation of emulator on test sets. Results are reported as $R^2$ values between GCAM and the emulator on its predictions (on the interpolated test set) and on input-output DGMS sensitivity (on the DGSM set), aggregated to region, year or quantity-level, and overall (no aggregation).}\label{tab:results}
    \begin{tabular}{lrrrr}
    & {\bf Region} & {\bf Year} & {\bf Quantity} & {\bf Overall}\\
    Predictions & 0.998 & 0.998 & 0.998 & 0.998 \\
    Sensitivity & 0.989 & 0.990 & 0.995 & 0.812 
    \end{tabular}
\end{table}

As summarized in Table~\ref{tab:results}, we analyze the performance of our emulator by comparing the output values to those of GCAM, as well as comparing the sensitivities of the emulator to those of GCAM. 
For the ``Predictions'' row of the table, 
we evaluate the ``Overall'' emulator performance on the interpolated test set by calculating the $R^2$ score for each of the 22,528 output values and report the median over these output values. This shows very high agreement with GCAM, with a median $R^2$ of 0.998. The results for Region, Year, and Quantity involve first aggregating targets over the other two dimensions (e.g., Region averages over Year and Quantity); $R^2$ is then computed for each the remaining outputs (44 if Quantity, 32 if Region, 16 if Year), and the median $R^2$ over these aggregated outputs is reported. This level of aggregation does not improve the already near-perfect overall $R^2$.

To further evaluate the quality of the emulator, we perform a Derivative-based Global Sensitivity Measure (DGSM) analysis  \cite{sobolDerivativeBasedGlobal2009}, as implemented in the SALib package \cite{Herman2017,Iwanaga2022}, on both our emulator and on GCAM. Specifically, we compare $S^\sigma_{ij}$ values defined as follows:
\begin{equation*}
    S^{\sigma}_{ij} = \frac{\sigma_{x_i}}{\sigma_{y_j}} S_{ij}, \mbox{ where }
S_{ij} = \mathbb{E} \left[ \left( \frac{\partial y_j}{\partial x_i} \right)^2 \right].
\label{eq:sigma-norm}
\end{equation*}
$S_{ij}$ is the $\nu$ value from \cite{sobolDerivativeBasedGlobal2009}, while $\sigma_{z}$ denotes the standard deviation of $z$. $S^\sigma_{ij}$ is a normalized version of $S_{ij}$; normalizing this way better captures the true effect of input $x_i$ on output $y_j$~\cite{saltelliGlobalSensitivityAnalysis2008}, given the wide range of magnitudes and units in GCAM inputs and outputs.
The sensitivity analysis uses the DGSM dataset, generated with the finite-diff sampling strategy; sensitivities are calculated by observing the effects of introducing small perturbations around each input parameter and seeing how each of the outputs respond. 
For the emulator and for GCAM, we calculate $S^\sigma_{ij}$ for all inputs $x_i$ and outputs $y_j$. 

The Overall result, summarized in Table~\ref{tab:results}, is the $R^2$ agreement between the $S^\sigma$ matrix for the emulator and the $S^\sigma$ matrix for GCAM. At 0.812, we observe good agreement between the emulator and GCAM with respect to the input-output sensitivities. For the Region, Year and Quantity breakdowns, we average the $S^\sigma$ matrices over disjoint subsets of output variables $j$, leaving only the specified dimension (e.g., the Quantity breakdown uses  $S^\sigma \in \mathbb{R}^{9 \times 44}$, having averaged sensitivities over Year and Region). Sensitivity agreement at this coarser resolution is very high, ranging from 0.989 for Region to 0.995 for Quantity. 

The input-output sensitivities, both of the emulator and GCAM, yield some interesting trends.
Most notably, there is a high normalized sensitivity to the \emph{energy} input factor for many of the outputs. This makes sense because this particular input variable affects the GDP and population assumptions, which past exploratory studies have also found to be the largest contributor to outputs \cite{dolanModelingEconomicEnvironmental2022a, kanyako2023compounding}. Predictably, we also see a strong sensitivity to the energy input factor among regions with large economies and high populations, such as China, India, and the USA.
Several of the output quantities stand out as highly sensitive to the inputs; in particular, \emph{electricity price} and many land sector outputs. Electricity price reflects the input drivers chosen for this ensemble, which experts selected specifically because they would affect energy prices from different technologies and therefore relative adoption of wind and solar. The land sector has been studied in past analyses \cite{dolanModelingEconomicEnvironmental2022a, kanyako2023compounding} showing that the inherently finite nature of land availability for feeding changing populations is often a key determinant of outcomes.  See Appendix~\ref{app:sensitivity} for additional information.

\section{Conclusion and Future Work}
\label{sec:conclusion}

We present in this paper a high-fidelity and computationally efficient emulator of GCAM using deep learning. In the process of doing so, we enriched the sampling strategy of inputs underpinning an existing exploration (in W2023) of the drivers of renewable energy deployment by 2050, relaxing 9 of 12 input variables from binary to continuous. This represents a particularly valuable addition to the past study that, by limiting exploration to binary choices for the input parameters, risked overlooking outcomes of interest associated with intermediate values.
We confirm that our emulator is highly accurate and that its sensitivities are consistent with GCAM's. 
In future work, we plan to explore the use of this high-fidelity emulator for searching over input space (e.g., to identify circumstances that minimize water scarcity) to steer the generation of large ensembles of GCAM, and to better understand GCAM itself.
Ultimately, we view this work as a bridge to a new era where large ensembles are still relevant, but their creation can be aided by machine learning to reduce the cost and complexity; future work to answer scientific questions around climate, energy, land and water systems can generate tailored ensembles in an iterative, emulator-in-the-loop manner.

\section{Acknowledgements}

This research was supported by the U.S. Department of Energy, Office of Science, as part of research in MultiSector Dynamics, Earth and Environmental System Modeling Program.  The Pacific Northwest National Laboratory is operated for DOE by Battelle Memorial Institute under contract DE-AC05-76RL01830. The views and opinions expressed in this paper are those of the authors alone.

\bibliographystyle{plain}
\bibliography{main}

\newpage

\appendix

\section{Inputs (Drivers)}
\label{app:inputs}
The 12 input variables are described in Table~\ref{tab:inputs}.

\begin{table}
    \caption{Inputs varied for each run of GCAM. Interpolated inputs in bold.}
    \label{tab:inputs}
    
    \begin{tabular}{llr}
        \toprule
            Input & Key & Description \\
        \midrule
            \textbf{Wind and Solar Backups} & back & Systems needed to backup wind and solar \\
            Bioenergy & bio & Tax on bioenergy \\
            \textbf{Carbon Capture} & ccs & Carbon storage resource cost \\
            Electrification & elec & Share of electricity in building, industry, and transportation \\
            Emissions & emiss & $CO_2$ emission constraints \\
            \textbf{Energy Demand} & energy & Energy Demand - GDP and population assumptions \\
            \textbf{Fossil Fuel Costs} & ff & Cost of crude oil, unconventional oil, natural gas, and coal \\
            \textbf{Nuclear Costs} & nuc & Capital overnight costs \\
            \textbf{Solar Storage Costs} & solarS & Solar storage capital overnight costs \\
            \textbf{Solar Tech Costs} & solarT & CSP and PV costs \\
            \textbf{Wind Storage Costs} & windS & Wind storage capital overnight costs \\
            \textbf{Wind Tech Costs} & windT & Wind and wind offshore capital overnight costs \\
        \bottomrule
    \end{tabular}
\end{table}

\section{Output Quantities}
\label{app:output-quantities}
Our 44 output quantities are described in Table~\ref{tab:output-details}.

\begin{table}
\centering
\begin{tabular}{lllll}
\toprule
resource & metric & sector & units & query name \\
\midrule
energy & demand\_electricity & building & EJ & elec\_consumption\_by\_demand\_sector \\
energy & demand\_electricity & industry & EJ & elec\_consumption\_by\_demand\_sector \\
energy & demand\_electricity & transport & EJ & elec\_consumption\_by\_demand\_sector \\
energy & demand\_fuel & building & EJ & final\_energy\_consumption\_by\_sector\_and\_fuel \\
energy & demand\_fuel & industry & EJ & final\_energy\_consumption\_by\_sector\_and\_fuel \\
energy & demand\_fuel & building & EJ & final\_energy\_consumption\_by\_sector\_and\_fuel \\
energy & demand\_fuel & industry & EJ & final\_energy\_consumption\_by\_sector\_and\_fuel \\
energy & demand\_fuel & transport & EJ & final\_energy\_consumption\_by\_sector\_and\_fuel \\
energy & price & coal & 1975\$/GJ & final\_energy\_prices \\
energy & price & electricity & 1975\$/GJ & final\_energy\_prices \\
energy & price & transport & 1975\$/GJ & final\_energy\_prices \\
energy & price & transport & 1975\$/GJ & final\_energy\_prices \\
energy & supply\_electricity & biomass & EJ & elec\_gen\_by\_subsector \\
energy & supply\_electricity & coal & EJ & elec\_gen\_by\_subsector \\
energy & supply\_electricity & gas & EJ & elec\_gen\_by\_subsector \\
energy & supply\_electricity & nuclear & EJ & elec\_gen\_by\_subsector \\
energy & supply\_electricity & oil & EJ & elec\_gen\_by\_subsector \\
energy & supply\_electricity & other & EJ & elec\_gen\_by\_subsector \\
energy & supply\_electricity & solar & EJ & elec\_gen\_by\_subsector \\
energy & supply\_electricity & wind & EJ & elec\_gen\_by\_subsector \\
energy & supply\_primary & biomass & EJ & primary\_energy\_consumption\_by\_region \\
energy & supply\_primary & coal & EJ & primary\_energy\_consumption\_by\_region \\
energy & supply\_primary & gas & EJ & primary\_energy\_consumption\_by\_region \\
energy & supply\_primary & nuclear & EJ & primary\_energy\_consumption\_by\_region \\
energy & supply\_primary & oil & EJ & primary\_energy\_consumption\_by\_region \\
energy & supply\_primary & other & EJ & primary\_energy\_consumption\_by\_region \\
energy & supply\_primary & solar & EJ & primary\_energy\_consumption\_by\_region \\
energy & supply\_primary & wind & EJ & primary\_energy\_consumption\_by\_region \\
land & allocation & biomass & thousand km2 & aggregated\_land\_allocation \\
land & allocation & forest & thousand km2 & aggregated\_land\_allocation \\
land & allocation & grass & thousand km2 & aggregated\_land\_allocation \\
land & allocation & other & thousand km2 & aggregated\_land\_allocation \\
land & allocation & pasture & thousand km2 & aggregated\_land\_allocation \\
land & demand & feed & Mt & demand\_balances\_by\_crop\_commodity \\
land & demand & food & Mt & demand\_balances\_by\_crop\_commodity \\
land & price & biomass & 1975\$/GJ & prices\_by\_sector \\
land & price & forest & 1975\$/m3 & prices\_by\_sector \\
land & production & biomass & EJ & ag\_production\_by\_crop\_type \\
land & production & forest & billion m3 & ag\_production\_by\_crop\_type \\
land & production & grass & Mt & ag\_production\_by\_crop\_type \\
land & production & other & Mt & ag\_production\_by\_crop\_type \\
land & production & pasture & Mt & ag\_production\_by\_crop\_type \\
water & demand & crops & km3 & water\_withdrawals\_by\_tech \\
water & demand & electricity & km3 & water\_withdrawals\_by\_tech \\
\bottomrule
\end{tabular}

\caption{GCAM outputs quantities with the associated GCAM selection query used to generated the outputs from the GCAM database.}
\label{tab:output-details}
\end{table}

\section{Input-Output Sensitivities}
\label{app:sensitivity}

\begin{figure}
    \includegraphics[width=\linewidth]{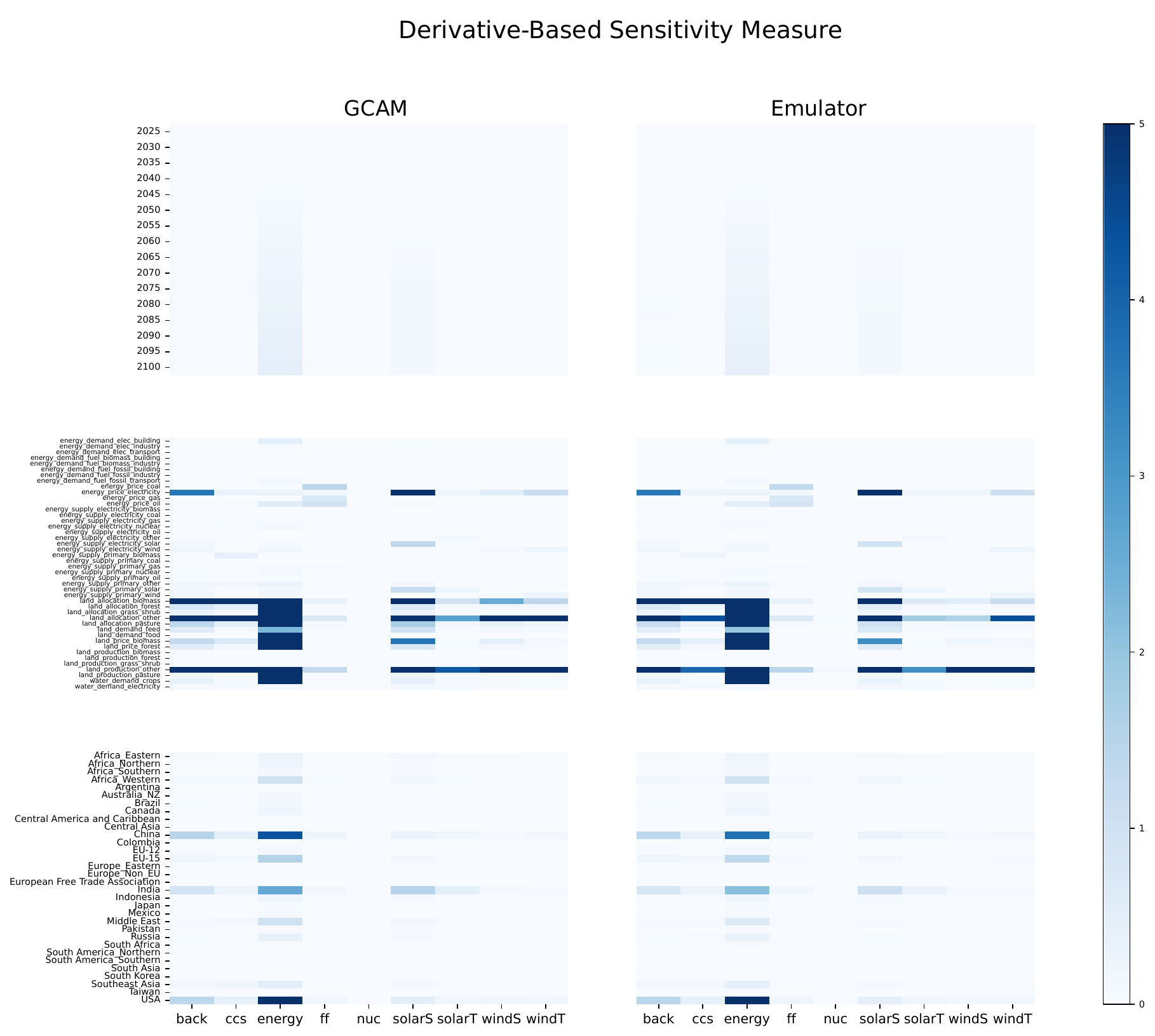}
    \caption{GCAM (left) vs. Emulator (right) local sensitivities of inputs vs Years (Top), Quantities (Middle), and Regions (Bottom).}
    \label{fig:sensitivity-heatmap}
\end{figure}

See Figure~\ref{fig:sensitivity-heatmap} for sensitivity values.

\end{document}